 \documentclass[final,3p,times,twocolumn]{elsarticle}
\usepackage{booktabs}
\usepackage{placeins}
\usepackage{graphicx}
\usepackage{amsmath}
\usepackage{color}

\usepackage{amssymb}
 \usepackage{lineno}

\journal{Energy and Buildings}

\begin{document}

\begin{frontmatter}

%% Title, authors and addresses

%% use the tnoteref command within \title for footnotes;
%% use the tnotetext command for theassociated footnote;
%% use the fnref command within \author or \address for footnotes;
%% use the fntext command for theassociated footnote;
%% use the corref command within \author for corresponding author footnotes;
%% use the cortext command for theassociated footnote;
%% use the ead command for the email address,
%% and the form \ead[url] for the home page:

%% \title{Title\tnoteref{labeL1}}

%% \tnotetext[label1]{}
%% \author{Name\corref{cor1}\fnref{label2}}
%% \ead{email address}
%% \ead[url]{home page}
%% \fntext[label2]{}
%% \cortext[cor1]{}
%% \address{Address\fnref{label3}}
%% \fntext[label3]{}

\title{Learning short-term past as predictor of human behavior in commercial buildings}

%% use optional labels to link authors explicitly to addresses:
%% \author[label1,label2]{}
%% \address[label1]{}
%% \address[label2]{}

\author[label1]{Romana Markovic\corref{cor1}}\ead{markovic@e3d.rwth-aachen.de}  \author[label1]{J\'{e}r\^{o}me Frisch} \author[label1]{Christoph van Treeck}

\address[label1]{E3D - Institute of Energy Efficiency and Sustainable Building, RWTH Aachen University, Mathieustr. 30, 52074 Aachen, Germany}
\cortext[cor1]{Corresponding author. Tel.: +49-241-80-25541 ; fax: +49-241-80-22030.}
\begin{abstract}

This paper addresses the question of identifying the time-window in short-term past from which the information regarding the future occupant's window opening actions and resulting window states in buildings can be predicted. The addressed sequence duration was in the range between 30 and 240 time-steps of indoor climate data, \textcolor{black}{where the applied temporal discretization was one minute}. For that purpose, a deep neural network is trained to predict the window states, where the input sequence duration is handled as an additional hyperparameter. Eventually, the relationship between the prediction accuracy and the time-lag of the predicted window state in future is analyzed. The results pointed out, that the optimal predictive performance was achieved for the case where 60 time-steps of the indoor climate data were used as input. Additionally, the results showed that very long sequences (120-240 time-steps) could be addressed efficiently, given the right hyperprameters. Hence, the use of the memory over previous hours of high resolution indoor climate data did not improve the predictive performance, when compared to the case where 30/60 minutes indoor sequences were used. The analysis of the prediction accuracy in the form of \textit{F1} score for the different time-lag of future window states dropped from 0.51 to 0.27, when shifting the prediction target from 10 to 60 minutes in future.

\end{abstract}

\begin{keyword}
%% keywords here, in the form: keyword \sep keyword
 neural networks  \sep stacked input vectors \sep sequence modelling \sep building automation systems \sep occupant behavior \sep window opening 
%% PACS codes here, in the form: \PACS code \sep code

%% MSC codes here, in the form: \MSC code \sep code
%% or \MSC[2008] code \sep code (2000 is the default)

\end{keyword}

\end{frontmatter}

% \linenumbers

%% main text

\section{Introduction}
% motivation for occupant behavior
Occupant behavior (OB) has been identified to be one of the principal factors influencing the energy consumption in commercial buildings \cite{wagner2017}, \cite{azar2014}, \cite{hong2013}, \cite{hong2016}. Due to that, developing an accurate model that predicts human actions would be beneficial for achieving higher indoor comfort or optimization of energy consumption.    
% Occupant behavior for building automation
Additionally, there have been a number of studies that addressed modeling the OB for its inclusion in building automation systems (BAS) \cite{dong2009}, \cite{yang2013}, \cite{majumdar2014}, \cite{peng2018}, \cite{mirakhorli2016}.
\\
% sequence modeling OB
According to the current research, OB in buildings is often defined as a discrete sequence in the temporal domain \cite{dong2009}, \cite{fristsh1990}, \cite{youngblood2007}, \cite{liao2012}. As such, not only is it necessary to identify which variables lead to occupants' actions, but also in which temporal range do the changes of variables in question occur, which motivated a  number of studies on time-series modeling of OB. Liao et al. \cite{liao2012} presented a probabilistic graphical model for depicting the time-series of occupancy data. Fritsch et al. \cite{fristsh1990} presented the model that generates time series of window opening angles with the same statistics as the measured openings for the heating period. Dong and Andrews \cite{dong2009} proposed occupancy pattern recognition using semi-Markov models. Youngbloot and Cook \cite{youngblood2007} introduced a hierarchical model for controlling the smart environment based on occupants' activities and concluded that learning algorithms built on Markov models experience performance issues when scaled to large problems. Wilke et al. \cite{wilke2013} modeled residential activities based on time-dependent probabilities to start activities and their corresponding duration distributions. Cali et al. \cite{cali2018} concluded that the Markov chain models have the clear advantage of properly including the time dependency of the process. Additionally, they pointed out that a logistic regression analysis --as an alternative to the proposed Markov model allows modeling based on more variables, such as the indicators of the indoor air quality.  
% deep learning
As already shown by Dont et al. \cite{dong2009}, a parallel can be drawn between speech recognition problems and OB modeling. Resultantly, the recent findings on efficient speech recognition modeling may be projected in the field of OB. For instance, graphical models and in particular Hidden Markov Models (HMM) were widely used for representing the temporal variability of speech \cite{hinton2012} since more than 30 years \cite{baker1986}. Although the HMMs are very efficient at solving a number of problems, their significant drawback is that probabilities need to be “hard coded”, which requires a new problem formulation for each task in question. On the other side, the deep neural networks do not require domain knowledge encoded into transition probabilities, since they can learn the feature mapping, given enough data that contains the underlying information. As a result, they outperformed graphical models in solving speech recognition tasks, usually by a large margin \cite{hinton2012}, \cite{graves2013}.

% deep learning for sequence modeling of OB

The problem of modeling sequences using neural networks in the field of OB and energy efficient buildings has already been addressed by a number of studies \cite{wang2018}, \cite{zhao2016}, \cite{coelho2017}, \cite{Kraipeerapun2017}, \cite{zhang2018}, \cite{wang2017}, \cite{li2017}, \cite{li2017a}. Coelho et al. \cite{coelho2017} proposed a deep learning method for forecasting the sequences of energy consumption in micro grids. Wang et al. \cite{wang2018} proposed a single layer recurrent neural network (RNN) for predicting the occupancy on the following time-steps, where the input consists of a WiFi signal, while Zhao et al. \cite{zhao2016} investigated multi-layer RNNs for predicting the occupancy based on physical measurements. Zhang et al. \cite{zhang2018} proposed HVAC control using deep reinforcement learning (DLR) with a feed-forward neural network. The simulation results pointed out, that the DLR approach lead to 15 \% lower heating energy consumption while perceiving thermal comfort. Wang et al. \cite{wang2017} presented a long-short term memory (LSTM) recurrent neural network for HVAC control. The results on the training and validation set pointed out, that the use of LSTM-RNNs could lead to improved energy efficiency and thermal comfort.
\\
However, there is little work available on the duration of the time-window from which the information regarding occupants' short-term future actions can be retrieved. For instance, Wang et al. \cite{wang2018} proposed modeling the occupants' presence using five past time-steps of a WiFi signal, where time-step duration was 30 seconds. Hence, the research question regarding the time-window of physical measurements of indoor climate required for algorithmically efficient formulations of predictive OB models remained open.
\\
% scope of this study
In the scope of this study, the time-window that contains the information about future OB actions is investigated. In particular, the following research questions are addressed:
\begin{itemize}
\item {what is the sequence duration in the short-term past, from which the information regarding the occupants' future action can be retrieved?}
\item {what is the relationship between the time-lag between the current time-step and the occupants' action, and the accuracy of the resulting predictive model?}
\item {the learning progress and problem formulation in case of stacked sequences of OB data will be analyzed.}
\end{itemize}
% method
For that purpose, a deep neural network is developed to predict the window states as the results of OB actions on future time-steps, where the input consists of indoor and outdoor climate information from the previous time-steps. The learning progress is analyzed, and eventually, the predictive performance was quantified.
\\
The remaining part of the paper is organized as follows: section 2 describes the  hypothesis and experimental setting; Section 3 includes the results – in particular, it addresses an optimal problem formulation, learning progress and resulting predictive performance. Eventually, the gained knowledge and limitations of this study are elaborated and summarized in Sections 4 and 5.

\section{Method}

\subsection{Experimental setting}
\begin{figure}[th!]
\centering
 \includegraphics[trim=1cm  8cm 0cm 0cm, width=0.45\textwidth]{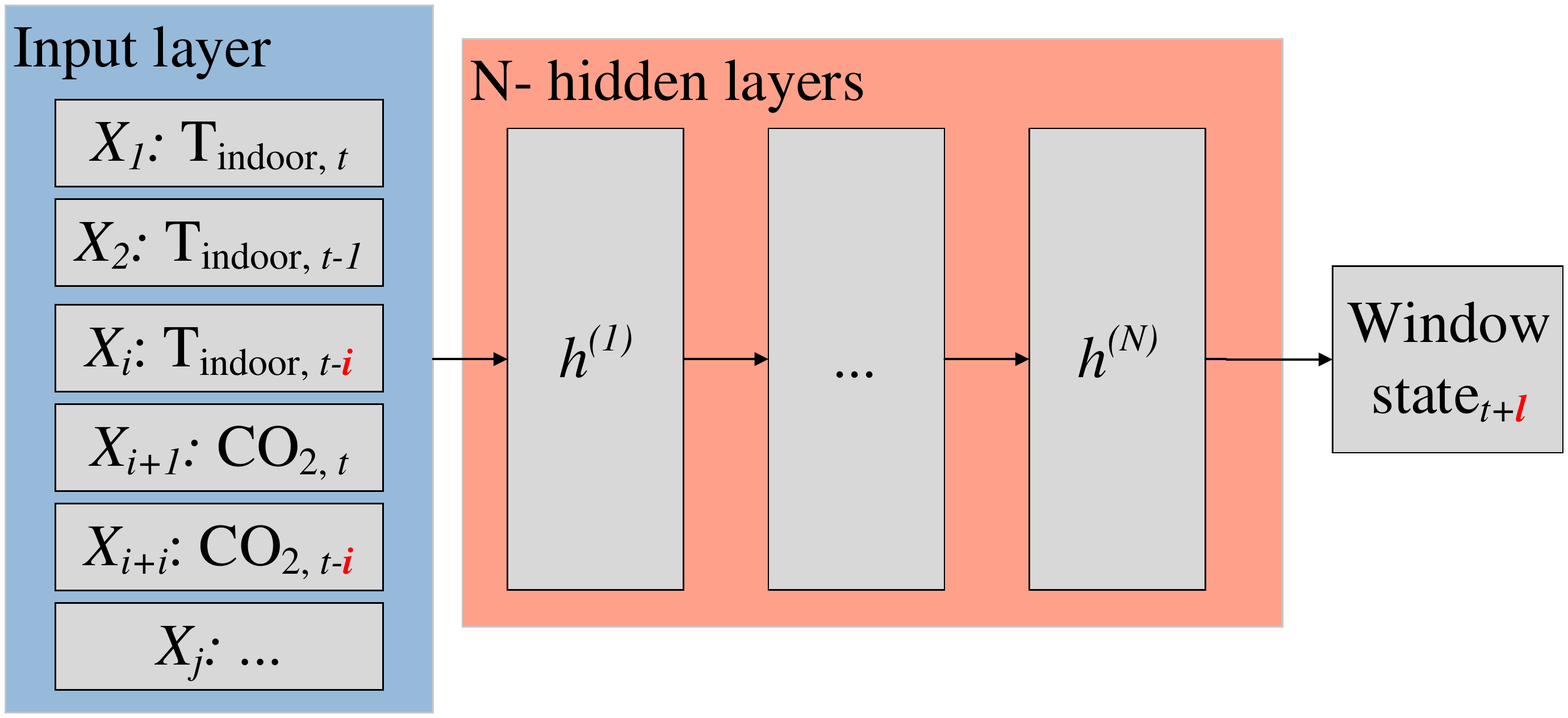}
\caption{Learning procedure overview. The sequences of measured indoor climate and the data from the last time-step \textit{t} are defined as the model inputs. The duration of the input sequences \textit{i} (red color symbol) is defined as additional hyperparameter. Eventually, the relationship between the predictive performance and the time-lag of prediction \textit{l} (red color symbol) is analyzed.  }
\label{fig:method}
\end{figure}

% give an overview of the proposed method
The starting hypothesis is, that OB actions are influenced by the changes in the indoor climate that occurred in the short-term past. For instance, an increase in indoor air temperature or presence of additional occupants (and resulting increase in indoor $CO_2$ levels) may carry the information regarding future window openings. For that purpose, the model input consisted of weather and climate data from the current time-steps, and additionally the indoor air temperature, indoor air humidity and $CO_2$ concentration measured at every time-steps from the short-term past. The short-term past was considered as time-window used as the input for the developed model. The first objective of this paper was to investigate the size of this time-window, by treating it as a hyperparameter during model training. An optimal time-window size was analyzed between 30 minutes and 4 hours in the 30-minutes steps. The input sequence consisted of between 30 and 240 \textcolor{black}{minute-wise} time-steps, which resulted in problem dimensionality between 101 (21 features at time-step \textit{t} \textcolor{black}{plus 3 sequences of 30 time-steps each}) and 741 (21 features at time-step \textit{t} \textcolor{black}{plus 3 sequences of 240 time-steps each}). 
\\
\textcolor{black}{Even though the RNNs are widely used neural networks for the sequence modeling, an alternative approach had to be applied for the case of addressed lenghts of OB data sequences.} Although there is no theoretical limit on maximal sequence duration for conventional RNNs \cite{mozer1995}, the experimental proofs showed that the successful back propagation through time (BPTT) is possible on up to 5000 time-steps in cases where the gradient clipping and regularization were applied \cite{pascanu2013}. However, based on the results from the recent empirical studies on modelling the computer vision and speech recognition problems (\cite{mclaughlin2016}, \cite{chung2014}), the maximal lengths of successfully modeled sequences were in the range between 15-30 time-steps. Hence, exploring OB sequences in the range of several hundreds of data points may be unfeasible using the above mentioned approaches. As a result, either a more complex learning method should to be applied (gated RNNs, echo networks or memory networks), or the problem may be simplified by "unrolling" the sequence through stacking the features for a feed forward neural network. \textcolor{black}{The latter concept found application in related fields (\cite{wu2015}, \cite{hamid2014}), and it was also applied in the scope of this study. Resultantly, the input time-window was stacked as additional features for a feed-forward neural network (Figure \ref{fig:method}), which resulted in relatively simple training, in comparison to RNNs.}

\FloatBarrier
\FloatBarrier

The time-step \textit{t} was defined as step on which the prediction of the future window states was made.  The input features were defined as vector \textit{X}, which consists of \textit{$X_{outdoor}$}- measured outdoor climate and temporal information at time-step \textit{t}, and the indoor climate information \textit{$X_{indoor}$} on all time-steps between \textit{t-i} and \textit{t}:
\begin{equation}
X= [X_{outdoor, t} ~~  X_{indoor, t}~~ X_{indoor, t-1} ~~ ...~~ X_{indoor, t-i} ].
\end{equation}
The window state was predicted for the future time-step, namely \textit{t+l}. Here, the window state was the output based on the input features \textit{X} and learned feature mapping \textit{g ($\cdot$)} through the hidden layers \textit{1} to \textit{N}: 

\begin{equation}
y_{t+l}=g^{(n)}(...g^{(1)}(W^{(1)} X)),
\end{equation}

where \textit{$W^{n}$}, \textit{n  $\in$ N} were the tuned weights. Since it was aimed to perceive the indoor climate information at a high resolution, the temporal discretization was fixed at 1 minute steps, while even the lower time resolution (which would hence result in information loss) may hardly be explored. An additional argument for keeping the high resolution inputs was that the problem dimensionality in question still remained low, when compared to alternative problems addressed by similar methods, such as simplified computer vision problems (for example 32 x 32 pixels \`{a} 3 channels).
\\
It was considered that not every minute in the time-window carried information regarding the future OB actions. Also, it was considered that the members of the sequence that carry information are highly variant. As a result, the aim was that a neural network could learn which time-steps are irrelevant. Resultantly, a suitable architecture should be given enough hidden neurons to learn the dependencies and a possibility to exclude surplus neurons and input features by setting them to zero. For that purpose, a regularization using \textit{L1} norm was introduced. Namely, \cite{ng2004} presented a theoretical proof that \textit{L1} results in efficient feature selection for the case similar to the problem addressed in this paper -- where data sets with a potentially large number of irrelevant features could be present. The \textit{L1} factor defines a subset of weights in the hidden layers to have zero as optimal value, which results in sparse weight matrices \cite{goodfellow2017}. As a result, \textit{L1} found a wide application as a feature selection method, which was also applied in the scope of this study. 
\\
Once the optimal input sequence duration was identified and the optimal model formulation was defined, the relationship between the time-lag of prediction in future and the predictive performance was explored. Here, the time-lag was defined as the time between the time-step of prediction \textit{t} and the time-step for which the prediction was made. The investigated cases include the time-lags between 10 and 60 minutes in the 10 minute-steps. For that purpose, the neural network with the hyperparameters as defined previously was retrained using the window state from  \textit{t+10} to \textit{t+60} as a target variable.

\subsection{Neural network training}
An optimal neural network architecture in terms of number of neurons per hidden layer and the number of hidden layers was investigated for each duration of input sequence. The number of hidden layers was investigated in the range between 2 and 9 hidden layers, while the optimal number of neurons per hidden layer was searched in range between 1 and 400. \textit{L1} regularization is introduced in order to avoid over-fitting, by setting some of the hidden neurons to zero. Here, the \textit{L1} penalty was tuned in range between 1e-1 and 1e-5. An optimal learning rate was searched in the range between 0.2 and 0.01. It is opted for an adaptive learning rate. The rectified linear units (ReLU) were used as activation functions. The number of training iterations for different duration of training sequences was set based on the changes in the learning process between 10k and 90k. The batch size was kept fixed at 128, due to cache memory constraints. Resultantly, between 2 and 18 epochs were required for each training process.
\subsection{Data set}
The data were collected on an university building in Aachen, Germany. The building has operable windows and mechanical ventilation available in all offices from the data set. For additional information on the used data, the reader is referred to \cite{markovic2018}, while the short overview of the  measured indoor climate conditions is presented in  Table \ref{tab:dataset}.
%\FloatBarrier
\begin{table}[ht!]

\centering
   \caption{Mean values of the indoor climate and window opening data for the used data set. }
     \begin{tabular}{lll}
\toprule
      \textbf{variable} & \textbf{unit} & \textbf{mean value}    \\
\bottomrule
\multicolumn{1}{l}{ indoor $CO_{2}$  } & ppm &514  \\
\multicolumn{1}{l}{$T_{indoor}$} & $^{\circ}$C  &  22.9\\
\multicolumn{1}{l}{indoor air humidity} & \%  &38.5  \\
\multicolumn{1}{l}{window opening actions} & 1/h & 1.07 \\
\multicolumn{1}{l}{proportion of window openings} & - & 0.07 \\
\bottomrule
\end{tabular}
\label{tab:dataset}
\end{table}  

 The training set consists of approximately 600k data points collected on three singe/double offices. The model suitability is optimized on the validation set that consists of 4 mio. data points. Eventually, the remaining 15 mio. data points were used for model evaluation.

\subsection{Computational environment}
The models were developed using the Tensorflow \cite{abadi2016} library for Python 3.5 and for Python 3.6. The hyperparameters were tuned using computational resources from the RWTH compute cluster (CPU only), the GPU-Cluster at RWTH Aachen (parallel access to max. 8 nodes \`{a} 2 GPUs, out of 57 available Nvidia Quadro 6000), and a single PC for mixed CPU/GPU computations (CPU- Intel Core i7-6900K (3.2 GHz) , GPU- Nvidia GeForce GTX 1080). All computational resources were running on Linux-based operating systems.

\section{Results}
\subsection{Input sequence duration}

\begin{figure}[th!]
\centering
 \includegraphics[trim=0cm  0cm 0cm 0cm, width=0.45\textwidth]{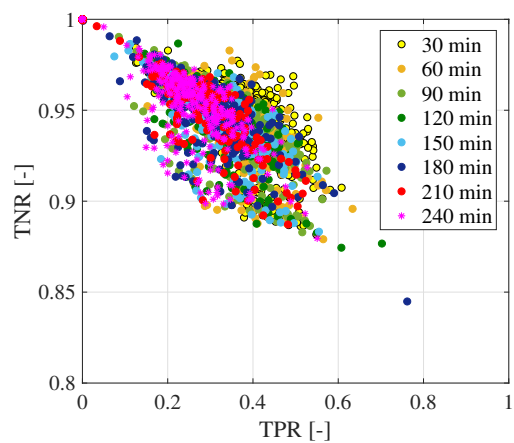} 
\caption{Qualitative representation of the relationship between the input sequence duration and validation performance. }
\label{fig:scatter_duration}
\end{figure}

A qualitative analysis of the impact of the input sequence duration on the model's predictive performance is conducted, with the aim to identify an optimal model formulation. Since the input sequence duration is handled as an additional hyperparameter, the accuracy as a function of the input duration is analyzed on the validation set. The resulting true positive rate (TPR) and true negative rate (TNR) pointed out, that the TPR rate was slightly lower for longer input sequences, when compared to 30 and 60 minutes input sequence durations (Figure \ref{fig:scatter_duration}). As a result, input sequence durations longer than 60 minutes did not lead to any performance improvement by incorporating "memory" regarding the occupants' actions and resulting window states over the previous 1-4 hours. Hence, a satisfying performance was also achieved for multiple hyperparameter combinations where longer input sequences were used. 
\FloatBarrier
\FloatBarrier
\textcolor{black}{
Based on the results of the hyperparameter search, a subset of suitable hyperparameter combinations is identified, and the training is repeated 50 times for each combination with the randomly chosen initial weights. The mean validation results over repeated trainings are summarized in Table \ref{tab:optimal_input}. 
}
The results pointed out, that no significant difference \textcolor{black}{could be observed in the validation performance}, in cases where the input sequences over 30 and 60 minutes were used. 
Accuracy was in the same range for each sequence duration, while the proportion of correctly identified opened windows dropped for 7 percent points, where the sequences longer than 120 minutes were used.

\begin{table}[h!]
\centering
   \caption{Validation results for 50 repeated model trainings using optimal hyperparameters for varied input sequence duration. }
  \color{black}   \begin{tabular}{lllll}
\toprule
\textbf{input sequence}	&	\textbf{ACC}	&	\textbf{TPR}	&	\textbf{TNR}	&	\textbf{F1}	\\    
\textbf{duration in}	&		&		&		&		\\
\textbf{minutes} &[-] &[-] & [-] & [-]\\
   \bottomrule
30	&	0.91	&	0.41	&	0.95	&	0.56	\\
60	&	0.90	&	0.41	&	0.94	&	0.55	\\
90	&	0.89	&	0.39	&	0.93	&	0.53	\\
120	&	0.89	&	0.39	&	0.93	&	0.54	\\
180	&	0.90	&	0.32	&	0.95	&	0.46	\\
240	&	0.88	&	0.32	&	0.92	&	0.45	\\
 \bottomrule

     \end{tabular}
   \label{tab:optimal_input}
\end{table}  
\FloatBarrier

\subsection{Optimal model formulation}

\begin{figure}[th!]
\centering
 \includegraphics[trim=1cm  0cm 1cm 2cm, width=0.45\textwidth]{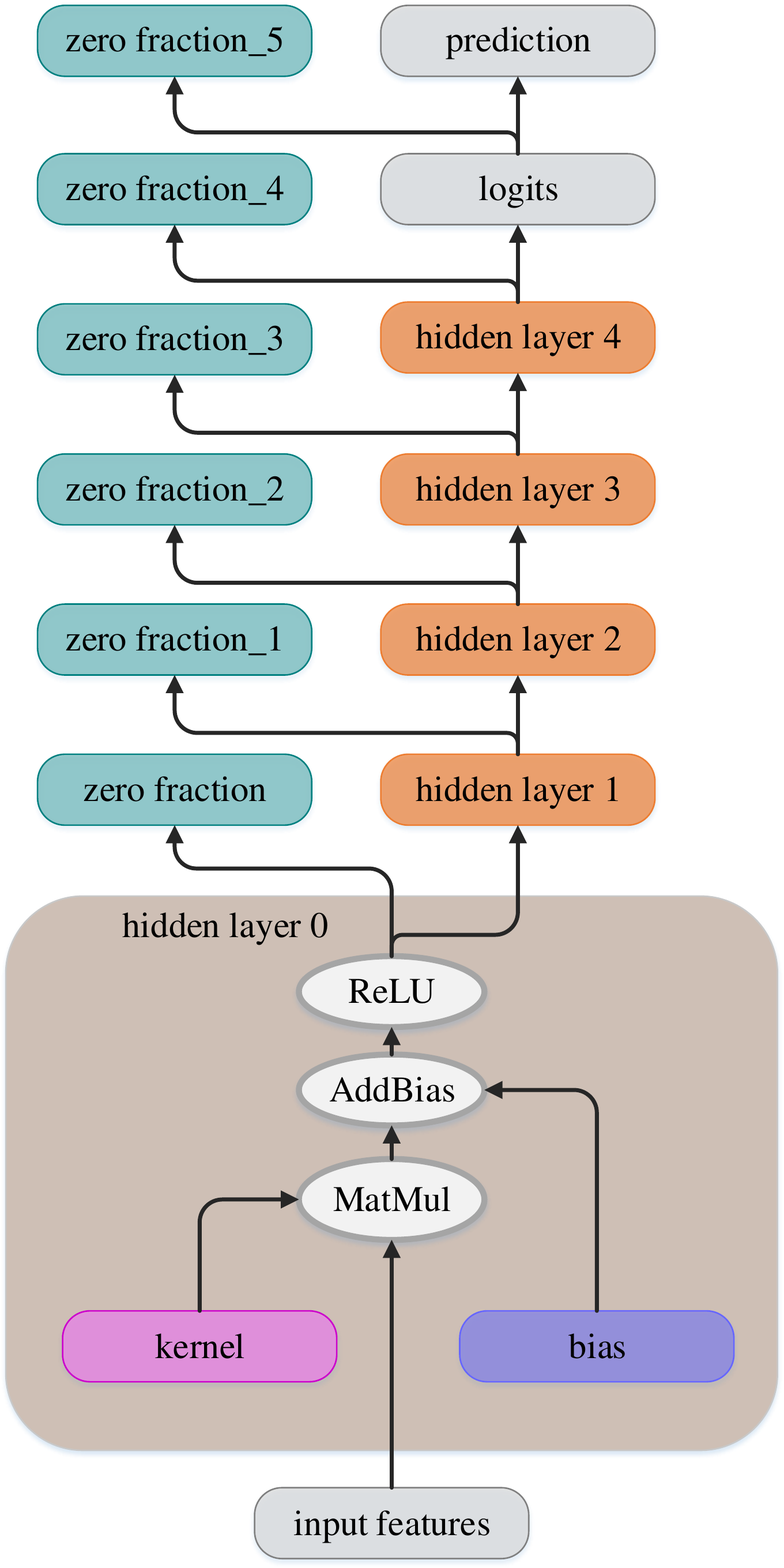}
\caption{Visualization of the developed neural network. \textcolor{black}{The graph notation corresponds to notation as used in Tensorboard \cite{abadi2016}.} }
\label{fig:graph}
\end{figure}

The developed main graph, and the structure of one hidden layer is visualized in Figure \ref{fig:graph}. The resulting network architecture consisted of five hidden layers with the following number of fully connected hidden neurons: 227-314-394-34-26. The \textit{L1} regularization coefficient was 0.01. The input sequence used was 60 minutes, where time-steps of 1 minute are selected. The model was trained for 50k iterative steps with the batch size of 128 data points (approximately 11 epochs) with the learning rate of 0.03.

\subsection{ Analysis of the learning progress}

\begin{figure}[th!]
\centering
 \includegraphics[trim=1cm  0cm 0cm 1cm, width=0.45\textwidth]{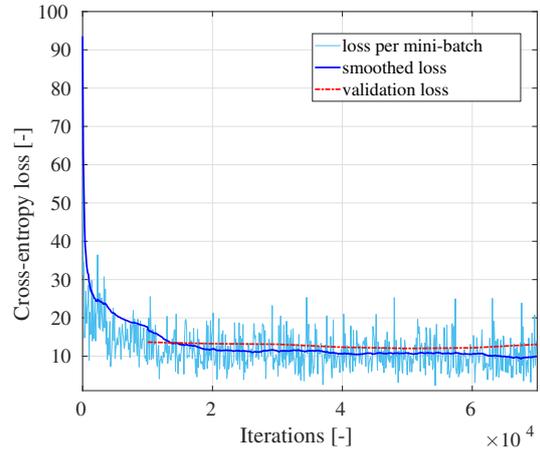}  
\caption{\textcolor{black}{Resulting training loss function (per mini-batch and smoothed) and validation loss.} }
\label{fig:loss}
\end{figure}

During the model training, the loss function formulated as the cross-entropy was minimized on the training set over multiple training epochs. Eventually, the training event at which the resulting validation error was minimal was identified as an optimal training duration. The training loss for each mini-batch and as a smoothed function is presented in Figure \ref{fig:loss}. Based on these results, the optimal number of training iterations was set to 50k with a learning rate of 0.03. Additionally, at the convergence of the loss function, the learning rate was reduced by factor 10 and eventually by factor 100, and additional 2x10k training iterations were conducted in order to reduce the variance in resulting predictive performance.

\FloatBarrier
\FloatBarrier

\FloatBarrier
\begin{figure}[th!]
\centering
 \includegraphics[trim=1cm  0cm 0cm 0cm, width=0.45\textwidth]{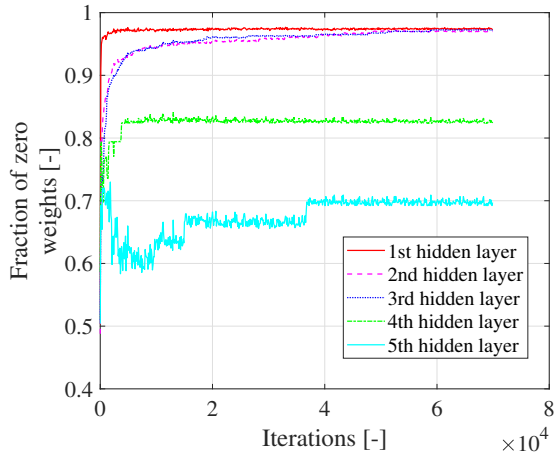}  
\caption{Fraction of the weights set to zero during model training. }
\label{fig:zero}
\end{figure}

\FloatBarrier
% regularization
The impact of the defined regularization on the size and complexity of the trained neural network is addressed by analyzing the tuned number of neurons per hidden layer and used weights in each layer. Figure \ref{fig:zero} presents the proportion of weights set to zero in each hidden layer. Namely, the input feature selection was conducted by setting 97 \% of the weights to the first layer to zero, which resulted in a sparse weight matrix. Additionally, the results of the learning progress resulted in later updates in the last hidden layer, which correspond to the improvements of accuracy presented in Figure \ref{fig:loss}. 
\FloatBarrier
\FloatBarrier

\begin{figure}[th!]
\centering
 \includegraphics[trim=1cm  0cm 0cm 0cm, width=0.45\textwidth]{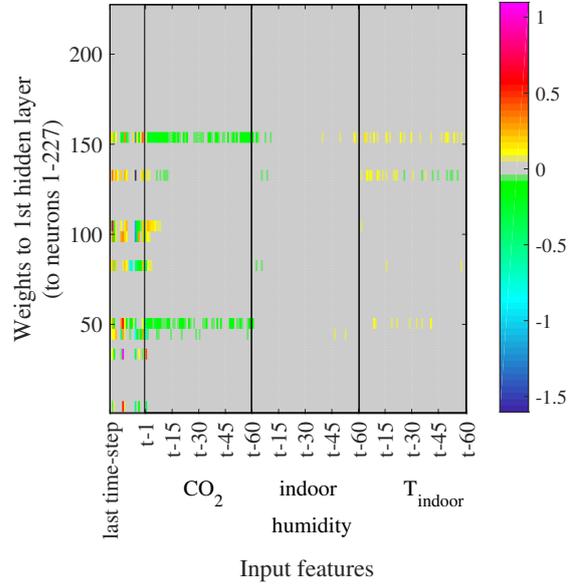}  \caption{Magnitude of the weights from the input layer (x-axis) to the neurons first hidden layer (y-axis). For the visualization purpose, the bars representing the weights of higher absolute magnitude (colored fields) are enlarged vertically by the factor 6.}
\label{fig:weights}
\end{figure}

\begin{figure}[th!]
\centering
 \includegraphics[trim=1cm  0cm 0cm 0cm, width=0.45\textwidth]{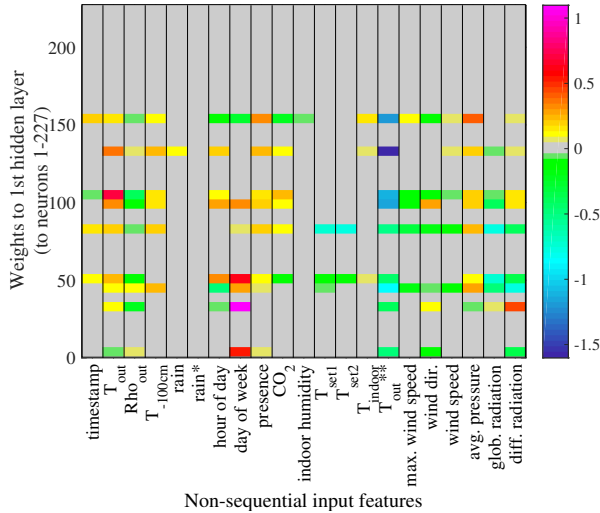}  %tuning_results.eps}
\caption{Magnitude of the weights from the input features that refer to non-sequential data, to the neurons in the first hidden layer. \textcolor{black}{For more detailed description of the input features, the reader is referred to \cite{markovic2018}.}}
\label{fig:weights_ts}
\end{figure}

In order to detect the relevant input features that were used to learn the mapping in hidden layers, the learned weights from the input layer to the neurons in the first hidden layer are presented in Figure \ref{fig:weights}. The weights from the input features sorted as indoor- and outdoor climate on the time-step of prediction and sequences of $CO_2$ measurements over last 60 minutes, indoor air humidity indoor and air temperature are presented on the x-axis, while the available neurons in the first hidden layers are listed out on y-axis.
\\
Based on the learned information and the tuned weights, the neurons may be classified in two groups. Firstly, a number of neurons learned the relevant information from the last time-step (colored bars on the left side of the diagram). A separate group of neurons learned the sequence of the changes in indoor climate together with the measurements from the last time-step. Namely, there were neurons responsible to learn densely represented $CO_2$ over time as the main prediction for window openings. Simultaneously, the sparse course of the indoor air temperature over last 60 minutes was identified as an additional predictor of window states. In addition, most of the weights in the matrix from indoor air humidity were set to zero. Resultantly, it may be interpreted that the neural network did not use a significant information from the changes in the indoor air humidity as a predictor of the future window states. 
\FloatBarrier
\FloatBarrier
The analysis of the learned weights from the non-sequential features to the first hidden layer showed that all input features besides the volume of precipitation contributed to the prediction of future window states (Figure \ref{fig:weights_ts})\footnote{Rain* referred to the volume of the rain droplets. $T_{out}$ refers to the air temperature at the weather station 1.3 km away from the building in question, while $T_{out}$** reefers to outdoor air temperature at the building site- there was up to 10 \% difference between the both values.}. Additionally, weights coming from the variables "day of the week" and "outdoor air temperature" had a larger magnitude, compared to alternative features. This may be interpreted that these variables have strong impact on the predicted window states. The sequence of the indoor air temperature over the past 60 minutes was learned to have impact on the future window states. However, the weights from the indoor air temperature on the previous step to the first hidden layer did not have large magnitude, when compared to to the alternative features. Based on these results, the trained neural network did not identify the indoor air temperature at the last time-steps as one of the main predictors of the future window states.

\FloatBarrier
\FloatBarrier

\subsection{What did the neural network learn?}
\begin{figure*}[t!]
\centering
 \includegraphics[trim=1cm  1cm 1cm 0cm, width=\textwidth]{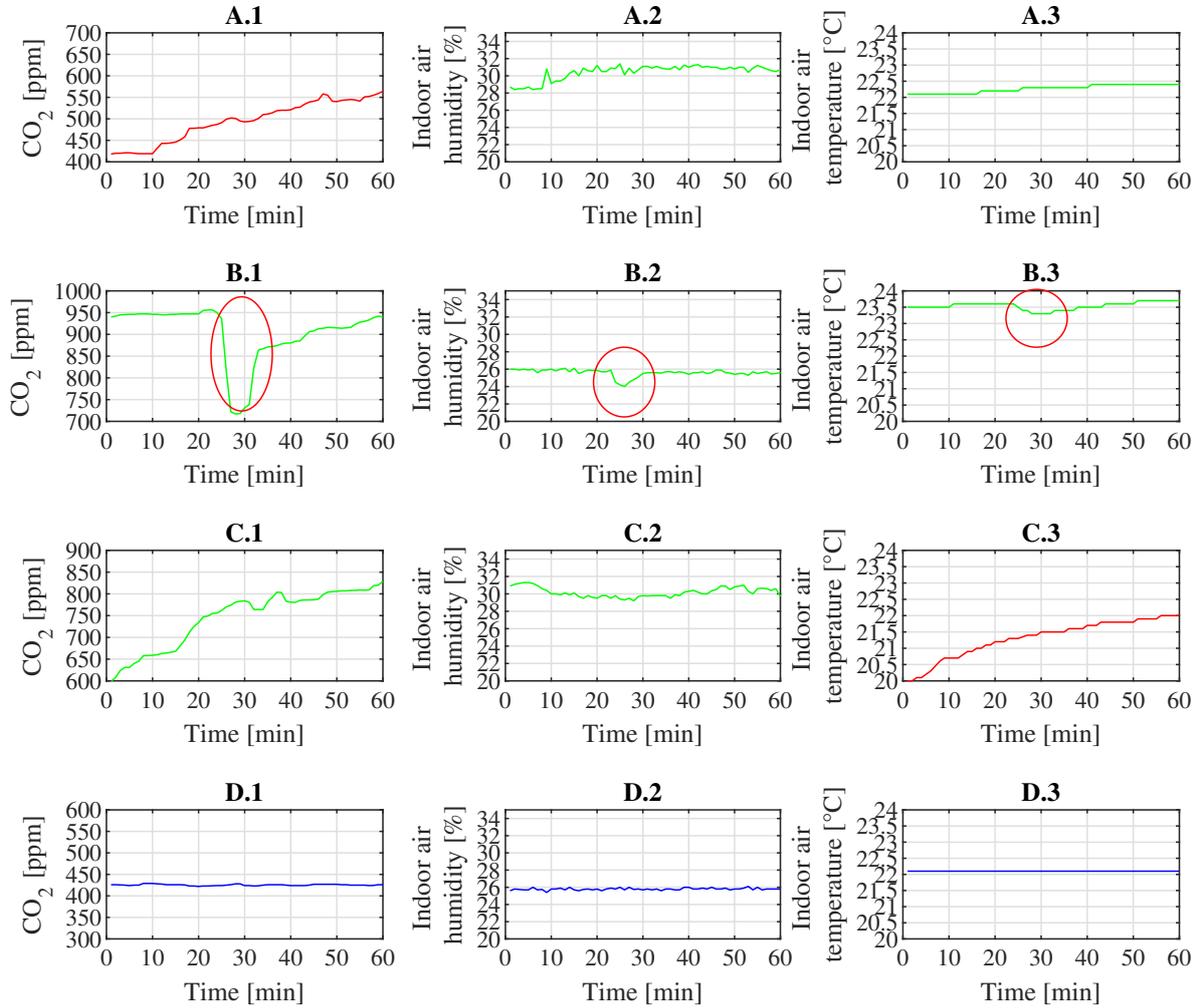}
\caption{Changes in the indoor climate prior over one hour, where the window states were correctly predicted to be opened. }
\label{fig:nn}
\end{figure*}

In the following Section, the domain knowledge will be applied to exemplary interpret the correctly predicted window states. The set of correctly identified window states consists of approximately 400k data points for the 10 minute time-lag and around 200k data points for the 60 minute time-lag. Since the elaboration on each of these data points would be unfeasible, 4 cases are extracted, where the network correctly predicted window states as open. The information that the trained neural network learned may be summarized into the following phenomena: 
\begin{itemize}
\item{events - the developed neural network learned that some occupants open the windows after the arrival (Figure \ref{fig:nn} (A.1-A.3))}
\item{past actions - the network could predict correctly that the window(s) will be opened again in 60 minutes, in case the short window opening occurred during the time-window used as the input sequence (Figure \ref{fig:nn} (B.1-B.3))}
\item{indoor climate and thermal comfort - the opened window state could be correctly predicted in case there was a steep rise in the indoor air temperature. This is of particular importance in the winter months, since a portion of the window opening caused by overheating could be correctly identified (Figure \ref{fig:nn} (C.1-C.3))}
\item{representations that could not be directly interpreted (Figure \ref{fig:nn} (D.1-D.3))}

\end{itemize}
\subsection{Predictive performance}

The results of 100 consecutive model trainings and evaluations on the test set using the same hyperparameters are presented in Table \ref{tab:acc_tp_tn}. The mean accuracy scored was 0.88, while the mean TPR and TNR were 0.37 and 0.92 respectively. The models' performance with respect to the data set's imbalance is quantified by the means of the \textit{F1} score, which had the mean value of 0.51. In addition to the performance mean values over repeated training with random initial weights, the extreme values (minimal and maximal performance for each metrics) and the 25\% and75\% quantiles were analyzed. Since these results showed low variance, the randomness of the weights initialization had no significant impact on the models' predictive performance.

\begin{table}[h!]
\centering
   \caption{Prediction performance of the investigated models for window opening. }
     \begin{tabular}{lllll}
\toprule
&\textbf{ACC} &\textbf{TPR}  & \textbf{TNR} & \textbf{F1}\\
&[-] &[-] & [-] & [-]\\
\toprule
min&	0.84	&	0.30	&	0.87	&	0.44	\\
25 \% quantile	&	0.87	&	0.35	&	0.91	&	0.49	\\
mean&	0.88	&	0.37	&	0.92	&	0.51	\\
median &	0.88	&	0.36	&	0.92	&	0.50	\\
75 \% quantile&	0.89	&	0.38	&	0.93	&	0.52	\\
max&	0.90	&	0.46	&	0.94	&	0.58	\\
 \bottomrule
     \end{tabular}
   \label{tab:acc_tp_tn}
\end{table}  
\FloatBarrier

The absolute performance results were analyzed using metrics proposed by \cite{mahdavi2016} and the results are summarized in Table \ref{tab:actions_duration}. The deviation between the observed and predicted proportion of time where windows were opened was 1 \%. The prediction results overestimated the number of opening actions per day by the factor 2.4. Resultantly, the durations of sequences where windows did not change the state (both open and close) were predicted to be lower, when compared to the measured durations of window openings and closings.
\FloatBarrier
\FloatBarrier
\begin{table}[h!]
\centering
\caption{Relative predictive performance. }

     \begin{tabular}{ccccc}
     \toprule
	 & \textbf{25 \% } 	&	\textbf{median	}&	\textbf{75 \%}  & \textbf{IQR\footnote{Distance between 25 \% quantile and 75 \% quantile.}} 		\\
\bottomrule
	 & \multicolumn{4}{l} {opening duration [hrs]}   \\
         \bottomrule	
observed &	0.15	&	0.51	&	1.63	&1.48\\
predicted& 0.05	&	0.18	&	0.67 & 0.62	\\
\bottomrule
& \multicolumn{4}{l} {closing duration [hrs]}   \\
\bottomrule
observed & 2.30	&	12.30	&	25.76& 23.46\\
predicted &0.07	&	0.52	&	11.15	& 11.08\\
\bottomrule
& \multicolumn{2}{l} {open state [-]} & \multicolumn{2}{l} {actions [1/d]}   \\
\bottomrule
observed & \multicolumn{2}{l} {0.07	}		&	\multicolumn{2}{l} {1.07}\\
predicted &\multicolumn{2}{l} {0.08	}		&	\multicolumn{2}{l} {2.10} \\
\bottomrule
     \end{tabular}
$^{2}$Distance between 25 \% quantile and 75 \% quantile.
   \label{tab:actions_duration}
\end{table}

The results showed, that 25 \% of the window opening actions could be correctly identified on the evaluation set. It is defined that the window opening action occurred, in case the transition from the closed to the open window state took place at least one time-step after the time-step of observation, and no later than at the time-step of the time-lag in future. Additionally, the prediction is handled as correct, in case the model predicted that the transition from the closed state, to the open window state will occur in the same time window.  

\subsection{Time-lag between observation and prediction}
\begin{figure}[th!]
\centering
 \includegraphics[trim=0cm  0cm 0cm 0cm, width=0.45\textwidth]{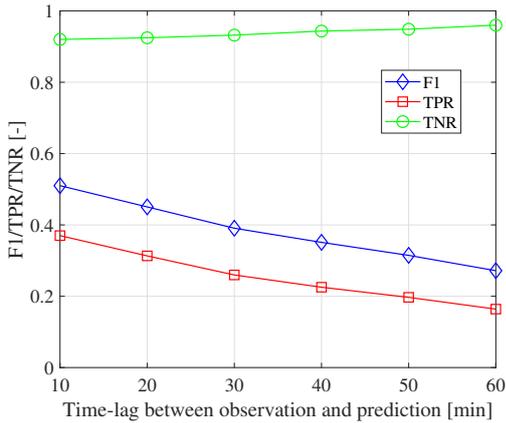}  \caption{Predictive performance for a varying time-lag of prediction.}
\label{fig:timelag}
\end{figure}

The optimal models with the 60-minute input sequences were re-trained for the case where the prediction target was between 20 and 60 minutes after the time of observation. Based on the training progress, the training was conducted for 20k more iterations (around 3 epochs using mini batches of 128 data points) for the cases where the time-lag was equal or grater than 30 minutes. Training with randomly chosen initial weights was repeated 100 times for each model and for each time-lag duration.
\\ 
The predictive performance for varied time-lag between the time of observation and the time of prediction for the optimal models with 60 minutes input sequences is presented in Figure \ref{fig:timelag}. The performance on the under-represented class (open windows) for both models was decreasing by 5-10\% for each 10 minutes of delay between the time of observation. Due to this decreased performance on the under-represented class, the TNR was slightly increasing in case of longer time-lag. 
\FloatBarrier
\FloatBarrier

\section{Discussion}
\textcolor{black}{
The duration of the model training was around 0.5 CPU seconds per 10k iterations, which resulted in 2-5 seconds for training a model with a single hyperparameter combination. However, the introduciton of the input sequence duration as an additional hyperparemeter resulted in an extensive grid search. In total, approximately 15k core hours of computations were conducted for the hyperparameter search, repeated training with random initial weights, analysis of the relationship between the time-lag and predictive accuracy and model evaluation. }
\\
% input sequence duration
The optimal input sequence duration has been identified to be 60 minutes, based on the predictive performance on the validation set for 100 consecutive model trainings with random initial weights. This resulted in a 201-dimensional problem formulation. Although the predictive accuracy in a similar range could be achieved in case where a longer input sequences were used, there was no performance improvement by incorporating the information regarding the indoor climate over periods between 60 and 240 minutes. The duration of sequences and the number of resulting time-steps used as model input are important for choosing suitable model architectures for further optimization and model development. Namely, the use of conventional RNN may result in an unstable model training due to exploding gradients due to long sequences that would be back-propagated through time (BPTT). As a result, the special case of RNN that include gated structures such as leaky units or long-short-term memory (LSTM) could be a suitable approach for further model development.
\\
\textcolor{black}{
Based on the trained weights from the input layer to the first hidden layer, the sequence of the indoor $CO_2$ measurements was identified as a significant predictor of future window states. Interestingly, it carried more information regarding the future OB actions, when compared to the current air temperature or the sequence of the indoor air temperature. Consequently, a high resolution in the temporal domain of the $CO_2$ measurements should be one of the inputs for the predictive models of OB in commercial buildings.}
\\
% learning 
\textcolor{black}{Additionally, the sparse weight matrix formulation resulted in a low number of activated neurons per hidden layer and resultantly computationaly efficient model evaluation}. Also, the regularization applied on input features resulted in a lower problem dimensionality. Namely, the indoor air humidity is identified to have low impact on predictions regarding future window open states. The results may be interpreted carefully in case of mechanically ventilated buildings, as it was the case for the building in question. Namely, the humidity could be constantly low in the monitored offices due to mechanical ventilation control strategy, so that no conclusion regarding the impact of humidity on the actions could be made. 
\\
This work referred to the analysis of the impact of the short-term past on the OB, but the knowledge gained could be used as a basis for analyzing the long-term impacts on OB such as acclimatization. In particular, a high temporal resolution (minute-wise) of indoor climate data would not be required for the measurements occurred more than 60 minutes in the past. Additionally, a similar implementation of the neural network could be applied for exploring longer time durations in lower resolution. For instance, introduction of multiple hidden layers could lead to accuracy improvement, and the similar regularization strategy could be applied.
\\
% energy saving potential
The purpose of this work is to explore empirically the input sequence duration required for modeling the human actions in commercial buildings. This is conducted on the sub-case of modeling the window states. This is motivated by the necessity to further explore the human behavior as a sequential problem in temporal domain. The direct applications of gained knowledge, which include models inclusion in BAS for energy saving and comfort optimization were not addressed in the scope of this work. However, they should be addressed in the future by the means of building performance simulation, laboratory studies or \textit{in-situ} studies on the buildings in operation.
\\
%  absolute performance
The absolute performance evaluation pointed out, that the neural network could accurately estimate the proportion of time where the window state was “open”, since the deviation between the measured and predicted open state was 1\%. The duration of sequences where windows state did not change were analyzed through the variables "opening duration" and "closing duration". The predictions underestimated the duration of both states (open/closed) where no change occurred. This may be caused by the lack of knowledge incorporated regarding the transition probabilities. Although the transition probabilities over the maximal input sequence duration of sixty minutes could be learned, the transition probabilites over longer time-series were not addressed. In order to achieve the end-to-end learning of transition probabilities over whole duration where the window state did not change, the used model must address the sequences' full duration. Since this may be inefficient using the proposed temporal discretization or proposed method, the introduction of the varied discretization in time domain could be a promising approach. Alternatively, models such as memory networks may be suitable for the problem in question.
\\
% limitations
The model was developed using data from an university office building in Germany. Validation and evaluation were conducted on a considerably large data set (training: 600k data points, validation: 4 mio. data points, evaluation: 15. mio data points). Since no experiments and additional evaluation were conducted using the data from different building types of different locations, no conclusions regarding the large scale model applicability across different building types or geographic locations could be raised. Due to that, independent model evaluations using additional data sets or round robin studies should be conducted in future.

\section{Conclusion}
\textcolor{black}
{
The goal of the presented work was to identify the time window in the short-term past where the information regarding the future OB actions was stored and to analyze the learning progress of deep learning driven predictive model of OB. The key findings may be summarized into the following:
}
\begin{itemize}
\item{\textcolor{black}
{addressing more than 60 minutes of high resolution indoor climate data did not lead to improvements in the models' predictive performance,
}}
\item{\textcolor{black}
{training of the deep neural network resulted in highly interpretable content, in terms of the weights from the input features, to the first hidden layer, }}
\item{\textcolor{black}
{a dense representation of sequential indoor $CO_2$ measurements was identified as one of the main predictors of window states, }}
\item{\textcolor{black}
{the predictive performance dropped with the increased forecasting horizon; resultantly, minute-wise predictions for more than 30 time-steps in the future could be made only with poor performance. }}

\end{itemize}

A significant contribution of the proposed method is learning end-to-end information regarding the drivers for future actions. Furthermore, the first hidden layer consisted of neurons “specialized” for phenomena that were interpretable and that were in accordance with the findings from the related studies based on domain knowledge. However, this can by no means be a replacement for research on understanding human actions in buildings. It may rather be considered as a supporting method, which allows researchers to extract the knowledge from a high volume of data, where conventional approaches may be unfeasible.

\section{Acknowledgements}

Simulations were performed with computing resources granted by RWTH Aachen University under project nova0015. The authors appreciate the financial support of this work by the German Federal Ministry of Economics and Energy (BMWi) as per resolution of the German Parliament under the funding code 03ET1289D. We thank Mark Wesseling and Davide Cali from EBC Institute, E.ON ERC at RWTH Aachen University for providing the monitoring data, as well as to the Physical Geography and Climatology Group for providing weather data.

\end{document}